\documentclass{wscpaperproc}
\usepackage{latexsym}
\usepackage{graphicx}
\usepackage{mathptmx}
\usepackage{enumerate}
\usepackage{enumitem}
\usepackage{amsmath}
\usepackage{amsfonts}
\usepackage{amssymb}
\usepackage{amsbsy}
\usepackage{amsthm}
\usepackage{cite}
\usepackage{algorithm, algorithmic}
\usepackage{subfigure}
\usepackage[utf8]{inputenc}

\usepackage[pdftex,colorlinks=true,urlcolor=blue,citecolor=black,anchorcolor=black,linkcolor=black]{hyperref}

\newtheoremstyle{wsc}
{3pt}
{3pt}
{}
{}
{\bf}
{}
{.5em}
{}

\theoremstyle{wsc}

\newcommand{\bz}{{\bf z}}
\newtheorem{Problem}{Problem}
\newtheorem{remark}{Remark}
\newtheorem{assumption}{Assumption}

\begin{document}

%
%

\pagestyle{fancyplain}

\thispagestyle{plain}
\firstPageHead{}

\chead{\fancyplain{}{\itshape Ghosh, Pal, Kumar, Ojha, Paranjape, Barat, and Khadilkar}}

\rhead{}
\cfoot{}
\renewcommand{\headrulewidth}{0pt} 

\makeatletter
\let\@internalcite\cite
\def\cite{\def\@citeseppen{-1000}%
    \def\@cite##1##2{(##1\if@tempswa , ##2\fi)}%
    \def\citeauthoryear##1##2##3{##1 ##3}\@internalcite}
\def\citeNP{\def\@citeseppen{-1000}%
    \def\@cite##1##2{##1\if@tempswa , ##2\fi}%
    \def\citeauthoryear##1##2##3{##1 ##3}\@internalcite}
\def\citeN{\def\@citeseppen{-1000}%
    \def\@cite##1##2{##1\if@tempswa, ##2)\else{}\fi}%
    \def\citeauthoryear##1##2##3{##1 (##3)}\@citedata}
\def\citeA{\def\@citeseppen{-1000}%
    \def\@cite##1##2{(##1\if@tempswa , ##2\fi)}%
    \def\citeauthoryear##1##2##3{##1}\@internalcite}
\def\citeANP{\def\@citeseppen{-1000}%
    \def\@cite##1##2{##1\if@tempswa , ##2\fi}%
    \def\citeauthoryear##1##2##3{##1}\@internalcite}
\def\shortcite{\def\@citeseppen{-1000}%
    \def\@cite##1##2{(##1\if@tempswa , ##2\fi)}%
    \def\citeauthoryear##1##2##3{##2 ##3}\@internalcite}
\def\shortciteNP{\def\@citeseppen{-1000}%
    \def\@cite##1##2{##1\if@tempswa , ##2\fi}%
    \def\citeauthoryear##1##2##3{##2 ##3}\@internalcite}
\def\shortciteN{\def\@citeseppen{-1000}%
    \def\@cite##1##2{##1\if@tempswa, ##2\else{}\fi}%
    \def\citeauthoryear##1##2##3{##2 (##3)}\@citedata}
\def\shortciteA{\def\@citeseppen{-1000}%
    \def\@cite##1##2{(##1\if@tempswa , ##2\fi)}%
    \def\citeauthoryear##1##2##3{##2}\@internalcite}
\def\shortciteANP{\def\@citeseppen{-1000}%
    \def\@cite##1##2{##1\if@tempswa , ##2\fi}%
    \def\citeauthoryear##1##2##3{##2}\@internalcite}
\def\citeyear{\def\@citeseppen{-1000}%
    \def\@cite##1##2{(##1\if@tempswa , ##2\fi)}%
    \def\citeauthoryear##1##2##3{##3}\@citedata}
\def\citeyearNP{\def\@citeseppen{-1000}%
    \def\@cite##1##2{##1\if@tempswa , ##2\fi}%
    \def\citeauthoryear##1##2##3{##3}\@citedata}
%
%
%
\def\@citedata{%
    \@ifnextchar [{\@tempswatrue\@citedatax}%
                  {\@tempswafalse\@citedatax[]}%
}

\def\@citedatax[#1]#2{%
\if@filesw\immediate\write\@auxout{\string\citation{#2}}\fi%
  \def\@citea{}\@cite{\@for\@citeb:=#2\do%
    {\@citea\def\@citea{, }\@ifundefined
       {b@\@citeb}{{\bf ?}%
       \@warning{Citation `\@citeb' on page \thepage \space undefined}}%
{\csname b@\@citeb\endcsname}}}{#1}}%

%
\def\@citex[#1]#2{%
\if@filesw\immediate\write\@auxout{\string\citation{#2}}\fi%
  \def\@citea{}\@cite{\@for\@citeb:=#2\do%
    {\@citea\def\@citea{; }\@ifundefined
       {b@\@citeb}{{\bf ?}%
       \@warning{Citation `\@citeb' on page \thepage \space undefined}}%
{\csname b@\@citeb\endcsname}}}{#1}}%

%
\def\@biblabel#1{}
\makeatother



\newdimen\bibindent
\bibindent=0.0em
\def\thebibliography#1{\section*{\refname}\list
   {}{\settowidth\labelwidth{[#1]}
   \leftmargin\parindent
   \itemindent -\parindent
   \listparindent \itemindent
   \itemsep 0pt
   \parsep 0pt}
   \def\newblock{}
   \sloppy
   \sfcode`\.=1000\relax}


\setlength{\baselineskip}{12.7pt}

\title{A SIMULATION DRIVEN OPTIMIZATION
ALGORITHM FOR SCHEDULING\\SORTING CENTER OPERATIONS}

\author{Supratim Ghosh \\
Aritra Pal\\
Prashant Kumar \\
Ankush Ojha\\
Aditya A. Paranjape \\
Souvik Barat \\
\vspace{12pt}
Harshad Khadilkar \\
Tata Consultancy Services Research \\
Galaxy Business Park, Sector - 62\\
Noida - 201309, UP, India\\
}

\maketitle

\section*{ABSTRACT}
Parcel sorting operations in logistics enterprises aim to achieve a high throughput of parcels through sorting centers. These sorting centers are composed of large circular conveyor belts on which incoming parcels are placed, with multiple arms known as chutes for sorting the parcels by destination, followed by packing into roller cages and loading onto outbound trucks. Modern sorting systems need to complement their hardware innovations with 
sophisticated algorithms and software to map destinations and workforce to specific chutes. While state of the art systems operate with fixed mappings, we propose an optimization approach that runs before every shift, and uses real-time forecast of destination demand and labor availability in order to maximize throughput. We use simulation to improve the performance and robustness of the optimization solution to stochasticity in the environment, through closed-loop tuning of the optimization parameters. 


\section{INTRODUCTION}
\label{sec:intro}

A sorting center is a sophisticated cyber-physical system comprising of infeeds, conveyor belts, different kinds of chutes, workforce, and outfeeds. For optimal operation, packages should move through the conveyor belt and the chutes in as little time as possible, with the additional goal of avoiding rejected parcels (ones that the automated system is unable to route, thus requiring human intervention). 

\subsection{Motivation for this work}

The parcel delivery industry was worth 500 billion USD in 2019 \cite{market_insight}, with further growth due to the pandemic in 2020. The large growth in an already high-demand industry has led to stress on existing infrastructure. Sorting centers are critical components of the parcel delivery logistics industry \cite{boysen2019automated}, since they are the points in the network where incoming parcels are aggregated and then segregated into onward destinations (see Section \ref{sec:desc} for details). The current practice in sorting terminal operations is largely manual and experience-driven. 
In order to maximize the throughput of these centers while retaining the existing infrastructure, one can attempt to optimize the sorting logic and workforce allocation to ensure a smooth flow of parcels. Such an algorithm is described in Section \ref{sec:method}. However, the stochasticity inherent in real-world operations cannot be completely captured in a formal optimization approach. Therefore, we use a high-fidelity simulation of sorting center operations (see Section \ref{sec:method}) to tune the constraints and parameters of the formulation, in order to ensure robustness (in the control theoretic sense, \cite{sastry2011adaptive}) while maximising throughput.

\subsection{Literature Review}
Optimization of a sorting center can be considered as a highly coupled, multi-level optimization problem. At the highest level, one can optimize the layout of the sorting center to maximize the efficiency \cite{werners2010}. At the next level (with a given sorting center layout), the optimization problems of interest typically involve (i) mapping chutes to destinations, (ii) mapping labor to chutes, and (iii) mapping individual parcels to chutes. This \textit{sort plan} is currently prepared in state-of-the-art applications once every few weeks based on expected demand \cite{jarrah2016destination,novoa2018flow}. However, to handle higher demand variability and with the benefit of earlier insight into demand through advanced sensing and forecasting, one can produce better solutions with smaller planning horizons.

At a fundamental level, the sorting problem bears resemblance to well-known 
queueing problems \cite{coff95}. To the best of our knowledge, theoretical tools used to study such ring networks have found little application in sorting, 
and this is very likely because they do not prescribe any particular optimal
algorithm. Instead, most of the literature on sorting has focused on linear programming and its variants. For instance, a robust planning approach to two-stage sorting operations is described in \cite{khir2021robust}. Their 
solution is robust to demand stochasticity from known uncertainty sets, but excludes labor assignment and assumes a fixed upper bound for the total number of parcels assigned to a chute. However, the real problem is dynamic because parcels are often loaded onto roller cages and cleared from the chute, 
thus creating space for additional parcels. It is worth noting that learning-based solutions have yet to be investigated in the literature for the sorting problem. However, reinforcement learning has been used for related problems such as inbound and outbound truck loading \cite{SHAHMARDAN2020106134}. A recent study \cite{kim2020sortation} uses reinforcement learning for the sortation problem, though this is defined as the task of routing parcels from `emitters' to `removers' through a grid. The definition is thus closer to a routing problem than the one considered in this study.

In this paper, we consider the combined problem of destination, parcel, and workforce mapping to individual chutes with the objective of maximising throughput with minimal disruptions (parcel rejections and chute blockages). Workforce mapping is an aspect that is often overlooked in literature. The problem involves assignment of $\ell$ near-identical laborers to $k$ chutes for a fixed time interval. Such problems can be solved in a static (as against sequential) setting using well-established techniques such as auctions \cite{vick61,aus00} or linear programming. However, the combination with destination and parcel mapping increases the complexity substantially. Given the high complexity and degree of coupling, we develop a simulation scheme for driving the optimization formulation towards realistic, implementable solutions. 
Among prior simulation studies, there are a few that use a software called FlexSim to simulation sorting center operations \cite{li2009flexsim,zhang2017study}. 
However, the flexibility of this software is not sufficient for the present work. In particular, we need to establish an iterative optimization-simulator loop for fine-tuning the solutions, as described in Section \ref{sec:method}.

\begin{figure}
    \centering
    \includegraphics[width=0.5\textwidth]{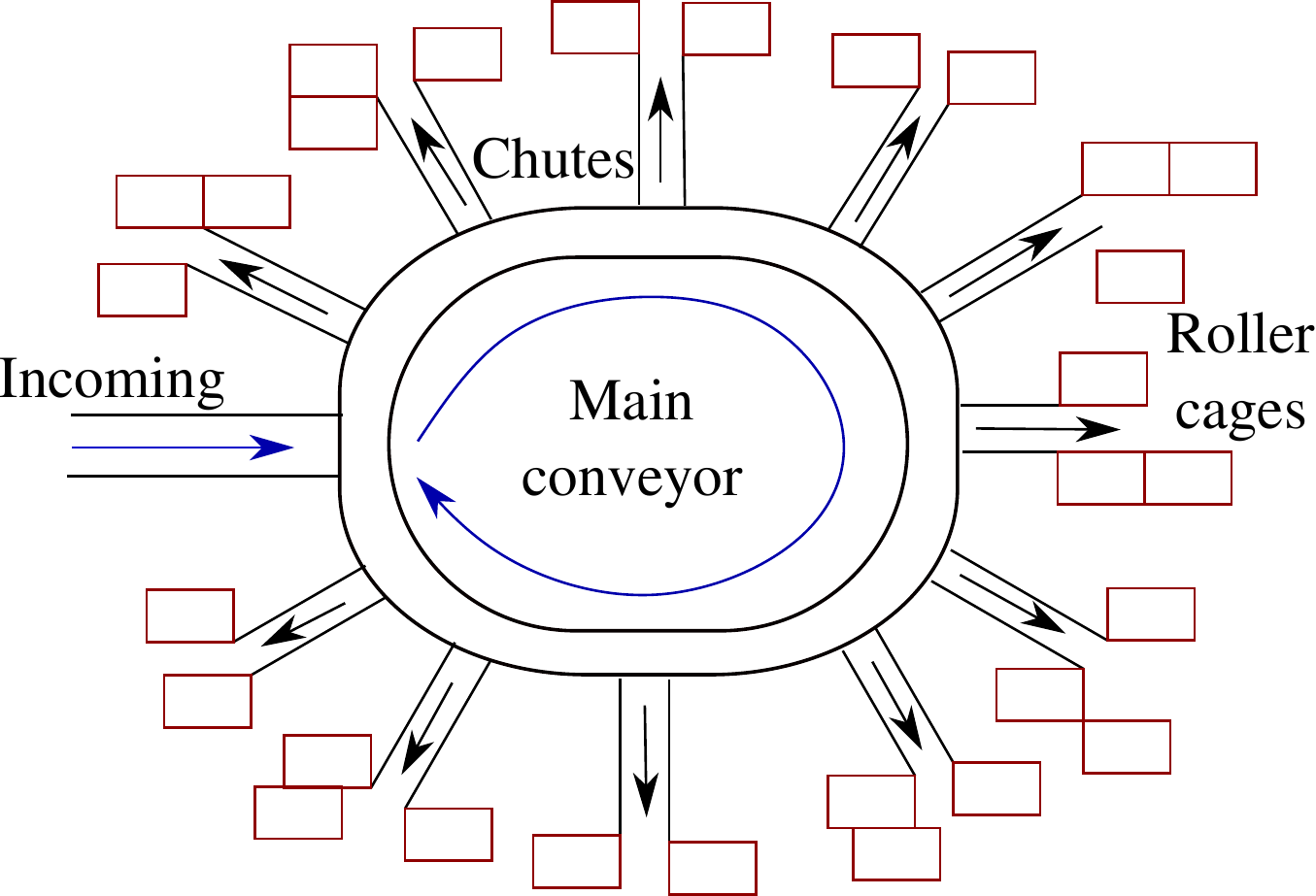}
    \caption{Schematic plan of a sorting center, showing the main oval shaped conveyor belt and outgoing chutes. Each chute is associated with a fixed number of roller cages, and each roller cage may be assigned to at most one destination, though multiple roller cages can serve the same destination.}
    \label{fig:schematic}
\end{figure}

\subsection{Contributions}
In the present work, we focus on a single-stage automated sorting center similar to the description in \cite{fedtke2017layout}, which discusses a number of design alternatives. A schematic layout is shown in Figure \ref{fig:schematic}, and a detailed problem description is provided in the next section. We solve the three individual sub-problems listed above (destination-chute mapping, labor allocation and parcel-chute mapping) separately, allowing us to handle multiple time scales inherent in the problem. 
Our contributions are:
\begin{enumerate}
    \item 
    Optimization algorithms for an offline destination$\rightarrow$chute plan and online parcel$\rightarrow$chute mapping;
    \item Integrating labor force allocation into the  sorting center optimization problem;
    \item 
    Using a simulation environment (digital twin) to (i) validate the optimization algorithms, and (ii) to iteratively improve the solution performance and robustness to stochasticity.
\end{enumerate}


\section{PROBLEM DESCRIPTION} \label{sec:desc}

In this section, we describe the specific sorting center layout assumed for the present work, and also describe the sorting problem in words. A mathematical treatment of the same is covered in the next section.

\subsection{System Layout} \label{sec:layout}

A schematic of the layout is shown in Figure~\ref{fig:schematic}, adapted from \cite{fedtke2017layout}. Parcels arrive at the sorting terminal via inbound trucks, and are processed in \emph{waves}. Each wave consists of a batch of incoming parcels for which a sort plan is computed. A complete shift ($8$ hours) typically consists of multiple waves (often around $10$). The objective of the system is to sort the incoming parcels by destination, and place them in roller-cages which are then loaded onto trucks for outbound transport.

The parcels to be processed in a wave are first processed by an OCR reader (not shown in the figure), which captures the intended destinations and the dimensions of the parcels. After this, the parcels enter the main conveyor. The system consists of multiple chutes, each connecting the conveyor (upper level) to a loading area with roller cages (lower level). Each parcel is diverted from the conveyor into the chute after allocation by an automated system for processing (placing into a roller-cage for onward transportation to its destination). We consider two types of chutes in this paper:
\begin{itemize}
    \item 
    \emph{Spiral Chutes:} These chutes consist of long metal tubes connecting the two levels. On the lower level, typically a human sorter or a robot picks up the parcel when it arrives and places it in the relevant roller cage according to destination. The number of destinations that this type of chute can handle is limited only by the number of roller cages it can hold.
    \item
    \emph{Direct chutes:} These chutes cater to exactly one location and are \emph{unsupervised}. In these chutes, the parcels are dropped from the upper level directly into roller-cages which are placed below (one cage per chute). Once, the cage is filled, it is immediately replaced by a new one. Direct chutes are usually reserved for destinations which have very high forecasted loads.
\end{itemize}
Each parcel on the main conveyor is either allocated a chute or is sent directly to the rejection chute (if no feasible chute is found). In addition, if a parcel is not able to enter a chute at the first attempt (due to chute blockage or dimension incompatibility), it can make a fixed number of re-attempts before being diverted to the rejection chute. The total number of parcels processed (packed into roller-cages and sent off for transportation) represents the \emph{throughput} of the sorting system and is measured in \emph{pph} (parcels per hour).


\subsection{The Sorting Problem}

The objective of our current work is to maximize the throughput of the sorting system subject to physical constraints on chute allocation, chute capacity, human resource allocation and capacity, and avoiding chute blockages. We solve the optimal sorting problem in two stages, \emph{planning} and \emph{execution}:
\begin{itemize}
    \item 
    \emph{Planning:} This stage takes the details of the projected load profile as input (total demand for each destination) and produces a matching between destinations and chutes \textit{for the entire shift} in such a way that the total number of parcels processed is maximized. Note that the matching is many-to-many: one chute can handle multiple destinations (equal to number of roller cages at that chute), and one destination can be mapped to multiple chutes. Special cases with constraints include (i) direct chutes, and (ii) layout constraints. Both are described in detail in the next section. Clearly, the forecast is noisy and other unexpected events may occur after planning. Thus a second online execution phase is necessary.
    \item
    \emph{Execution:} Once the planning is done and the parcels start arriving, the execution stage starts. This stage comprises of three parts: (i) scanning a parcel to get its dimensions and intended destination, (ii) allocating the parcel to a specific chute, and (iii) final processing after parcel comes out of the chute using roller-cages and trucks. The optimization algorithm performs the second part; i.e., it allocates specific chutes to each parcel in the upcoming wave (typically composed of thousands of parcels). This is in contrast to the planning stage, which only mapped destinations to a subset of chutes. The input to the algorithm is the destination for each parcel and the designated chutes for the destination as defined by the planning stage. Other constraints such as maximum parcel holding capacity of each chute are also considered, and so is workforce allocation.
\end{itemize}
The importance of simulation in this process is emphasised by the multi-step nature of the planning and execution phases. Since the planning phase is restricted to partial demand knowledge and labor availability (forecasts), an open loop system could potentially result in sub-optimal operations. Furthermore, the optimization constraints (defined later) are designed to handle worst-case scenarios in terms of chute blockage and parcel rejection, and may result in overly conservative plans. In order to avoid both these problems, we use the simulator in the planning phase to roll out multiple demand scenarios and to adjust the constraint thresholds for the optimization problem. The next section covers this aspect in detail.

\begin{figure}
    \centering
    \includegraphics[scale=0.3685]{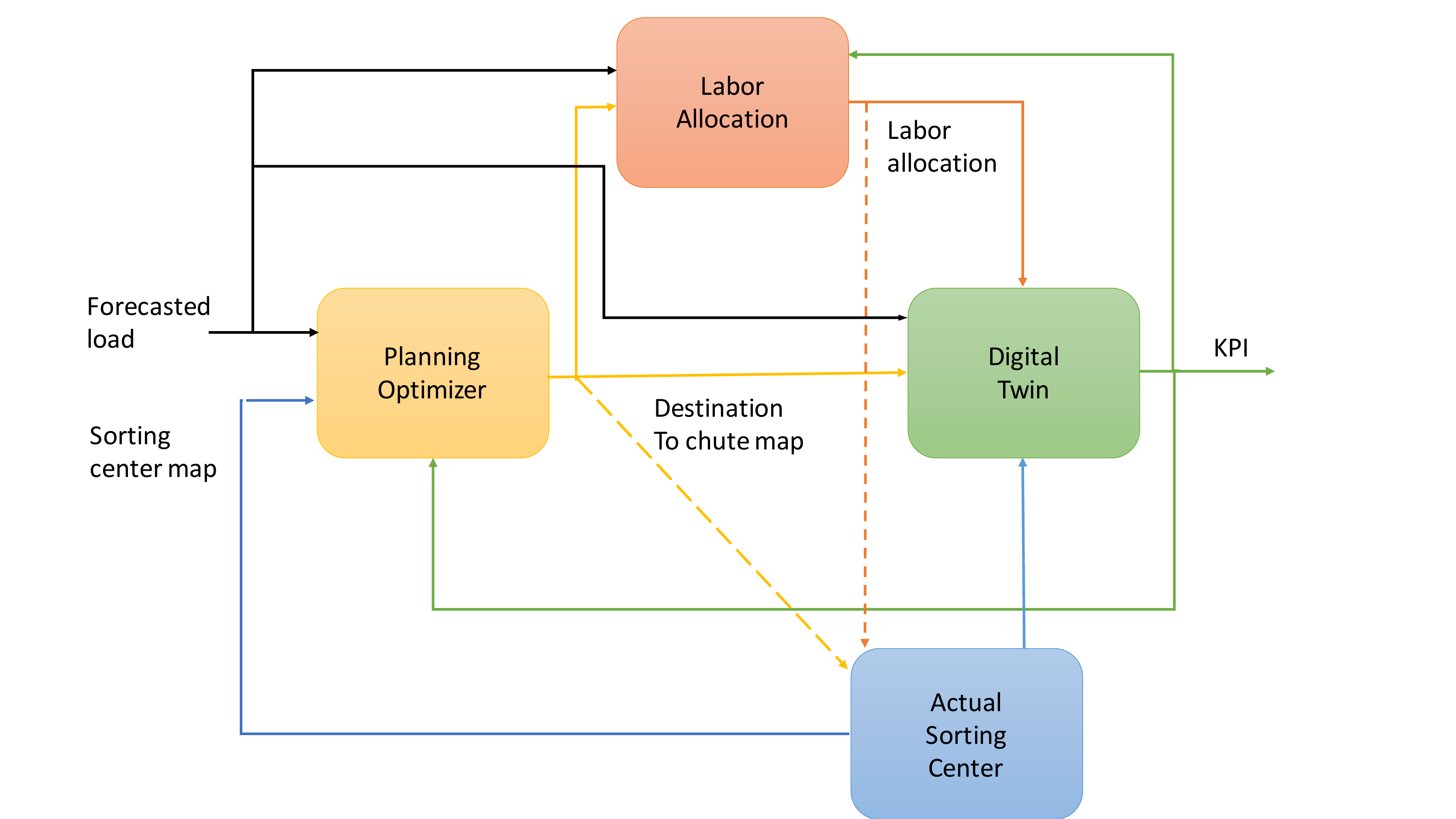}
    \caption{An overview of our methodology. Dashed lines indicate information flow.}
    \label{fig:method}
\end{figure}

 \section{METHODOLOGY} \label{sec:method}

We rely on optimization and its improvements via feedback from a simulation environment (called the digital twin) to solve the optimal sorting problem as shown in Figure~\ref{fig:method}. We first present and implement the optimization formulation OPTSORT (a mixed-integer linear program) for the two stages described in Section~\ref{sec:desc}. Following this we validate our results using digital twin (described later in this section). We then investigate how some of the constraints in this formulation can be gradually relaxed with the goal of improving the throughput without creating any chute blockages and violation of other physical constraints. 

\subsection{Optimization Algorithm OPTSORT}

\subsubsection{Planning}
The objective of this stage is to produce a destination-chute matching based on the projected load so that a maximum of such parcels are processed. Let the set $\mathcal{C}\triangleq\{C_1,\dots,C_k\}$ denote the $k$ chutes available in the sorting center and $\mathcal{D}\triangleq\{D_1,\dots,D_n\}$ denote the $n$ destinations to which the parcels in the projected load will be sent. Correspondingly, let $B_i$ denote the number of parcels in the projected load that will be sent to destination $D_i$. We will denote the total shift length by $T$ and the time taken to process one parcel at a chute $C_j$ by $t_j$. The quantity $t_j$ depends upon a variety of factors such as type of chute (Direct/Spiral); number of human resources manning the chute; and efficiency of those resources. We will present a separate algorithm of human resource allocation later in the paper.
We need to define the following variables:
\begin{align*}
X_{ij}\in\{0,1\} &= \left\{
                \begin{array}{l}
                1 \text{ if destination }D_i\text{ is allocated to chute }C_j ,\\
                0, \text{ otherwise,}
                \end{array}
                \right. \\
Y_{ij} \in \mathbb{Z}_{\geq 0}&= 
                \text{number of parcels of destination }D_i \text{ allocated to chute } C_j,
\end{align*}
Without specific limits, a many-to-many matching between destinations and chutes will be created. To avoid the situation where a destination is distributed to \emph{too} many chutes and vice versa, we impose certain limits: $M_i$ (respectively, $N_j$) will denote the maximum number of chutes (resp., destinations) the destination $D_i$ (resp., chute $C_j$) can be matched to, for $i\in\{1,\dots,n\}$ (resp., for $j\in\{1,\dots,k\}$); 
Note that $M_i$ and $N_j$ can be adjusted based on designer preferences. The following MILP produces the destination-chute matching with the goal of maximizing the number of processed parcels from projected load:


\begin{align}
\max &\sum_{i=1}^n\sum_{j=1}^k X_{ij}\nonumber\\
\text{subject to }&1\leq \sum_{j=1}^{k} X_{ij}\leq M_i \text{ for all }i\label{eq:destmax}\\
&\sum_{i=1}^{n} X_{ij}\leq N_j \quad\text{ for all }j\label{eq:chutemax}\\
&\sum_{j=1}^{k} Y_{ij}\leq B_j\quad \text{ for all }i\label{eq:destcap}\\
&\sum_{i=1}^{n} Y_{ij}\leq \left\lfloor\frac{T}{t_j}\right\rfloor\quad \text{ for all }j\label{eq:chutecap}\\
&X_{ij}\leq Y_{ij}\leq \mathcal{M}X_{ij}\text{ for all }i,j,\label{eq:xyreln}
\end{align}
where $i\in\{1,\dots,n\}$, $j\in\{1,\dots,k\}$, and $\mathcal{M}$ is a sufficiently large positive integer. Constraint \eqref{eq:destmax} ensures that every destination is matched to at least one chute; and that destination $D_i$ is not matched to more than $M_i$ chutes (for every $i\in\{1,\dots,n\}$). A similar bound with respect to destinations per chute is imposed in \eqref{eq:chutemax}. Constraint \eqref{eq:destcap} ensures that no more than $B_i$ parcels can be processed for destination $D_i$ while \eqref{eq:chutecap} guarantees that the number of parcels allocated to chute $C_j$ do not exceed the maximum number that can be processed at $C_j$. The final constraint \eqref{eq:xyreln} ensures that if destination $D_i$ is matched to chute $C_j$, then at least one parcel intended for destination $D_i$ is processed at $C_j$. This MILP formulation is suitable for situations where we have only Spiral chutes with no layout restrictions. Furthermore, we have two special cases:
\begin{itemize}
    \item 
    \emph{Direct chutes:} We follow a simple rule for Direct chutes: allocate the destinations with highest projected loads to Direct chutes. The remaining chutes and destinations can be matched according to the above formulation.
    \item
    \emph{Restricted Layouts:}  These situations involve additional restrictions of destination-chute matching wherein a destination can only be matched to a (strict) subset of $\mathcal{C}$. Such a context arises for e.g.; when the departure bays for the trucks departing to destinations are categorized by directions or zones and we are trying to minimize the distance travelled by the roller-cages from chutes to departing trucks. To impose this restriction, we introduce a new parameter $A_{ij}\in\{0,1\}$ such that $A_{ij}=1$ if and only if, destination $D_i$ \emph{can be} allocated to chute $C_j$ and add an additional constraint $X_{ij}\leq A_{ij}$ for all $i\in\{1,\dots,n\}$ and $j\in\{1,\dots,k\}$ in the formulation.
\end{itemize}
\subsubsection{Execution}

After the shift planning is completed with the creation of a destination-chute matching; real-time processing of parcels start. The parcels are processed in waves (mini-batches) which typically have to be processed in about an hour. The inputs to OPTSORT in this phase are: (i) parcel-destination data for the wave (i.e., which parcel goes to which destination); (ii) list of time-stamps at which the parcels enter the main conveyor; and (iii) destination-chute matching from the planning phase.
We now propose the MILP for creating the parcel-to-chute allocation based on these inputs: Let the set $\mathcal{P}=\{P_1,\dots,P_N\}$ denote the $N$ parcels arriving as input to the current wave. Let the intended destination for parcel $P_m$ be $D_{i_m}$ with $m\in\{1,\dots,N\}$ and $i_m\in\{1,\dots,n\}$. Using the destination-chute matching from the planning phase, we then create parameters $Q_{mj}\in\{0,1\}$ with $m\in\{1,\dots,N\}$ and $j\in\{1,\dots,k\}$ where $Q_{mj}=1$ if and only if $X_{i_mj}=1$; i.e.; the intended destination for $P_m$ is matched to chute $C_j$ and hence, parcel $P_m$ \emph{can be} allocated to chute $C_j$ (the exact chute allocation is provided by OPTSORT).

\begin{assumption}\label{ass:system}
We make the following assumptions about the operation of the system: 
(i) the conveyor moves at constant speed; i.e., the time taken by any parcel to reach the opening of a particular chute from the OCR reader is fixed and known; (ii) the parcels in a wave arrive sequentially in increasing order of their ids;  and (iii) for each destination matched to a chute; at least one roller-cage is open all the time which is immediately replaced after it is filled.
\end{assumption}

 Let $\tau_m$ for $m\in\{1,\dots,N\}$ be the time-stamp at which parcel $P_m$ enters the sorting system; i.e., it crosses the OCR reader. Also, let $\bar{\tau}_j$ with $j\in\{1,\dots,k\}$ denotes the time taken for a (any) parcel to reach the opening of chute $C_j$ from the OCR reader. Correspondingly, we can define $\tau_{mj}=\tau_{m}+\bar{\tau}_j$ as the time-stamp at which parcel $P_m$ is present at the opening of chute $C_j$. Assumption~\ref{ass:system} guarantees that  $\tau_{m_1}\leq \tau_{m_2}$ and consequently, $\tau_{m_1j}\leq\tau_{m_2j}$ whenever $m_1<m_2$ for all $m_1,m_2\in\{1,\dots,N\}$ and for all $j\in\{1,\dots,k\}$. Let $T_w$ denote the total time within which a wave needs to be processed. For a chute $C_j$, we define $t_j$ as the time taken to process one parcel, $L_j$ as the bound on maximum number of parcels that can be processed within one wave (if such a bound exists based on labor laws etc.), and $\mathcal{C}_j$ as the carrying capacity of chute $C_j$. Essentially, $\mathcal{C}_j$ denotes the total number of parcels that can stay inside chute $C_j$ without blocking it. 
 
 Note that the quantity $t_j$ depends upon the number of human resources allocated to a chute and their efficiency. We present an algorithm for human resource allocation later in this section.  Armed with these constants, we now define the binary variable $\bar{P}_{mj}$ such that $\bar{P}_{mj}=1$ if and only if parcel $P_m$ is allocated to chute $C_j$ for $m\in\{1,\dots,N\}$ and $j\in\{1,\dots,k\}$. A parcel is considered \emph{processed} if it is allocated to a chute. The objective of OPTSORT is to maximize the total number of parcels processed; i.e., to
 \begin{equation}\label{eq:obj2}
     \text{maximize }\sum_{m=1}^N\sum_{j=1}^k\bar{P}_{mj}.
 \end{equation}
The following constraint ensures that parcel $P_m$ \emph{is} allocated to chute $C_j$ only if it \emph{can be} allocated to $C_j$:
\begin{equation}\label{eq:canbe}
    \bar{P}_{mj}\leq Q_{mj} \quad\forall m\in\{1,\dots,N\} \text{ and }\forall j\in\{1,\dots,k\}.
\end{equation}
Next we have the bound on total number of parcels that can be allocated to chute $C_j$ within a single wave:
\begin{equation}\label{eq:chutebound}
    \sum_{m=1}^N \bar{P}_{mj}\leq \min\left\{\left\lfloor\frac{T_w}{t_j}\right\rfloor,L_j\right\} \quad\forall j\in\{1,\dots,k\}
\end{equation}
Following this, we have the constraint which ensures that one parcel is allocated to at most one chute:
\begin{equation}\label{eq:parcelbound}
    \sum_{j=1}^k \bar{P}_{mj}\leq 1 \quad\forall k\in\{1,\dots,N\}
\end{equation}
Finally, we formulate a constraint to ensure that the chutes are never blocked. This is done by guaranteeing that a situation where a spate of parcels enter a chute before the earlier ones are processed \emph{never} happens:
\begin{equation}\label{eq:chuteblock}
\sum_{m\colon \tau_{mj}\in[r,r+\mathcal{C}_jt_j]}\bar{P}_{mj}\leq \bar{\mathcal{C}}_j \quad\forall r\in\{1,\dots,T-\mathcal{C}_jt_j\}; \;\forall j\in\{1,\dots,k\}.
\end{equation}
The initial value of $\bar{\mathcal{C}}_j$ is set equal to $\mathcal{C}_j$ initially, which is refined based on feedback from the digital twin. This optimization formulation allocates parcels to specific chutes. In situations where we may have extra human resources available or we want to allocate more (or less) than one person per chute depending upon actual load; a resource allocation algorithm needs to be implemented which is described next.
\subsection{Allocation of Human Employees to Chutes}
The problem of assigning humans to chutes is equivalent to
mapping $n$ almost identical objects (e.g., resources with
figures of merit drawn from a normal distribution with a small covariance) to 
$p$ identical bins (e.g., tasks). We note one particular challenge in the context of sorting centers, namely that some chutes (e.g., those that receive heavy or
bulky packages) may require at least two humans for operation.

Let $z_{ij} \geq 0$ denote the penalty associated with $C_i$ when it is assigned $j-1$ workers which can be interpreted as a 
function of the parcels that will be {\em unprocessed} if $j-1$ workers are 
assigned to $C_i$.
\begin{remark}
It is possible to use more complex functions of the 
number of unprocessed parcels; for instance, the square
of the unprocessed fraction (taken with respect to the 
capacity) can be used to relax the penalty if the
number of unprocessed parcels is relatively small.
We assume the following for all $i,\, j$:
\end{remark}
\begin{eqnarray}
\nonumber \textrm{Non-increasing penalty:}& & z_{ij} \geq z_{i,j+1} \\
\nonumber \textrm{Non-increasing marginal penalty:} & & z_{ij} - z_{i,j+1} \geq  z_{i,j+1} - z_{i,j+2}~{\rm
except}~{\rm if}~z_{ij} = z_{i,j+1} > 0
\end{eqnarray}
\begin{remark}
The second condition helps satisfy a necessary condition for local optimality, namely that neighboring allocations yield a higher
penalty than the one obtained from the algorithm. 
We have that $z_{ij} = z_{i\,j+1} > 0$ when the addition of an 
exactly one extra individual to the $i^{\rm th}$ chute does not
help reduce the penalty. This can happen for chutes wherein each
package needs at least two handlers. Finally, we define the matrix of penalties and and its columns as 
\end{remark}
\begin{equation}
Z = [\bz_{1},\dots,\bz_{n+5}]^\top = \{z_{ij}\},~~Z \in \mathcal{R}^{p \times n+5} 
\label{eq:Zmat}
\end{equation}
The reason for adding four extra columns is purely
to support the notation employed in our algorithm
(see Algorithm~\ref{algo:1}). As such, all entries of 
those columns can be trivially set to $0$.
The assignment problem can be cast formally as follows.
\begin{Problem}
Suppose we are given a matrix $Z \in \mathbb{R}^{p \times n+1}$, where $z_{ij} \geq z_{i,j+1} \geq 0$ for all $i,\,j$. We wish to determine an assignment $(i,\,\sigma(i))$ to solve the following problem: 
\begin{equation}
\min_{\sigma}\sum_{i = 1}^p z_{i\sigma(i)}~~{\rm s.t.}~
\sum_{i = 1}^n \sigma(i) = p
\label{eq:J}
\end{equation}  
\end{Problem}
\begin{algorithm}[htb]
\caption{Assignment of workers to chutes}
\label{algo:1}
\begin{algorithmic}
\STATE Initialize: $\sigma(i) = 0~\forall\,i$; $k_{\rm last} = \emptyset$ 
(previous assignment)
\STATE Initialize: $C_0 = \bz_{1}$ (current assignment); $C_1 = \bz_{2}$ (assign
one individual to a chute); $C_2 = \bz_{3}$
\STATE Initialize $a = \sum_i \sigma(i)$
\STATE Define $P_2 = \{k \in [1,p]~|~C_0(k) = C_1(k) > 0\}$; !chutes with packages needing two handlers
\WHILE {$a < p$ AND $\max_k C_0(k) > 0$}
\STATE $k = {\rm arg}\max_k (C_0(k) - C_1(k))$
\STATE $\delta_1 = C_0(k) - C_1(k)$
 !most profitable reduction of penalty \\
\STATE ! Check if two people should be assigned to a single
 chute in $P_2$
\IF {$a \in (1, n-2]$ AND $P_2 \neq \emptyset$}
\STATE $\delta_0 = C_0(k_{\rm last}) - C_1(k_{\rm last})$
! last profitable allocation
\STATE $\delta_c = \max_{j \in P_2} (C_0(j) - C_2(j))$ ! reduction in penalty from allocating to a $P_2$ chute
\IF {$\delta_c > \delta_0 + \delta_1$}
\STATE $k = {\rm arg}{\max} (\delta_c)$
$\sigma(k) \leftarrow \sigma(k) + 2$, $a \leftarrow a + 2$
\STATE $C_0(k) = C_2(k)$; $C_1(k)= \bz_{\sigma(k)+3}(k)$; 
$C_2(k)= \bz_{\sigma(k)+4}(k)$
\ELSE
\STATE $\sigma(k) \leftarrow \sigma(k) + 1$, $a \leftarrow a + 1$
\STATE $C_1(k) = C_2(k)$; $C_2(k) = C_3(k)$; $C_3(k)= \bz_{\sigma(k)+3}(k)$
\ENDIF
\ENDIF
\STATE $k_{\rm last} = k$
\ENDWHILE
\STATE {\bf Output}: $\sigma(1:p)$
\end{algorithmic}
\end{algorithm}

The assignment algorithm is listed in Algorithm~\ref{algo:1}. 
It assigns individuals sequentially, aiming to achieve the largest possible reduction in penalties at each step. We now describe the simulation environment, an \emph{enterprise digital twin} of a sorting center terminal which will act as both a medium of validation and provide feedback to OPTSORT for improvement.

\subsection{Digital Twin} 

\begin{figure*}[htb]
    \centering
    \includegraphics[width=0.7\textwidth]{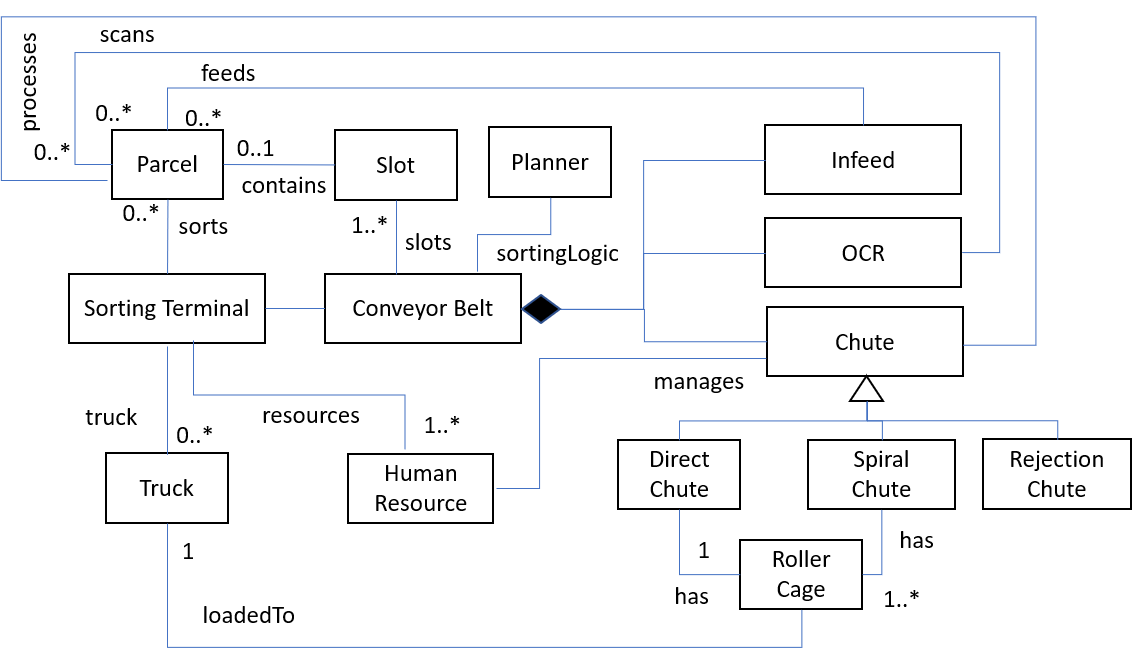}
    \caption{A schematic representation of sorting digital twin.}
    \label{fig:entities}
\end{figure*}


We adopt a pragmatic modelling \& simulation methodology for evaluating optimization outcomes using a close-to-real environment. Principally, we use a purposive hi-fidelity digital twin (DT) of a sorting terminal that captures all relevant entities, behavioural aspects, inherent uncertainties and pragmatic considerations. A schematic representation of a typical sorting terminal digital twin is shown using a class diagram in Figure~\ref{fig:entities}. It has three parts: a cyber-physical system, a \emph{planner} (in this case it is the optimizer) and a set of \emph{human resources}. From a structural perspective, the conveyor belt has multiple \emph{slots} (to hold and control parcel movements) and a set of sub-systems: \emph{infeeds}, \emph{OCRs}, and \emph{chutes} (Direct/Spiral/Rejection). Once the parcels are processed at Direct or Spiral chutes; they are put into roller cages which are finally loaded onto different \emph{trucks} for final dispatch. A three-step process to construct a faithful DT is described below:

    \smallskip
     \textbf{Construction \& contextualization}: We constructed a configurable sorting terminal digital twin by adopting a bottom-up actor-based \cite{agha1997foundation} modelling paradigm and an actor-based specification language - ESL \cite{clark2017esl}. Behaviours and interaction patterns of the entities shown in Figure~\ref{fig:entities} are captured as \emph{actors}. For example, all parcels, conveyor belts, slots, chutes, and resources are realised as actors. Entity behaviours, such as movement of conveyor-belts and slots, tilting of slots at specific chute, and resource emptying parcels from Spiral chute to roller cages, are realized as actor behaviours. All practical considerations and inherent uncertainties of the entities, such as varying efficiency of resources and arrival time of parcels, are encoded as probabilistic actor behaviours.
    
    Appropriate configuration parameters are introduced to represent a specific sorting terminal of a logistic company that includes specific number of conveyor belts, slots for each conveyor belt, number and position of in-feed, OCRs, and chutes, chute sizes, and resource efficiencies.

    \smallskip
    \textbf{Validation}: We use operational validity technique as suggested by \cite{sargent2010verification} to establish the faithfulness of a contextualized digital twin. Essentially, parcels' arrival details and specific sorting logic of a set of historical days are fed into contextualized digital twin and their simulation results are compared with real outcome to establish faithfulness.      
    
    \smallskip
    \textbf{What-if simulation}: A validated digital twin is used as an environment to evaluate the efficacy of optimization under different practical considerations. In this step, a sequence of parcels and an optimized sorting logic are fed into the digital twin and specific KPIs (i.e., key performance indicators), such as throughput, chute blockage, number of rejections, etc. are observed through simulation as shown in Figure~\ref{fig:entities}. Simulation of the digital twin involves executing the probabilistic micro-behaviour of all entities and observing the emergent macro-behaviour to understand the efficacy of a sorting logic under different parcel loads. A sorting logic can be evaluated for different parcel loads by feeding different streams of parcels to the digital twin. Similarly, different sorting laws for the same parcel load and a specific sorting logic with different parcel loads under various practical considerations (e.g., limited personnel/hour, workers with less productivity, mechanical failures) can be tested using our method shown in Figure~\ref{fig:method}.         
    

\section{RESULTS AND DISCUSSION}

In this section, we present the results of our experiments. The key ideas of our experiments are: 
\begin{itemize}
    \item
    Implement and validate the results produced by OPTSORT;
    \item 
    If the chute blockage constraint \eqref{eq:chuteblock} turns out to be conservative, can we gradually relax it to improve throughput and KPIs (defined later) without adding chute blockages?; and
    \item
    Study the robustness of OPTSORT to variations in human efficiency.
\end{itemize}

We classify our experiments into three groups: i) no restrictions on destination-chute matching (\emph{unrestricted}); ii) restrictions on which destination can be mapped to which chutes (restricted layouts); and iii) Direct chutes combined with restricted layout situations. We assume the following parameters for experiments: $n=300$, $k=30$ with $M_i=5$, and $N_j=15$ for every $i\in\{1,\dots,n\}$ and $j\in\{1,\dots,k\}$. Every wave is to be processed in $3000$ seconds. For the planning stage, we assume a load that is distributed randomly for the $300$ destinations to be processed within $30,000$ seconds ($10$ waves). Moreover, for the constrained layout scenario we assume the following matching restrictions: $D_1-D_{60}\to C_1-C_6$; $D_{61}-D_{120}\to C_7-C_{12}$; $D_{121}-D_{180} \to C_{13}-C_{18}$; $D_{181}-D_{240} \to C_{19}-C_{24}$; and $D_{241}-D_{300}\to C_{25}-C_{30}$.
Furthermore, we assume that there is one person manning one chute who is able to process $120$ parcels/hour; i.e., time taken to process a parcel at a chute is equal to $t_j=30$ seconds for every $j\in\{1,\dots,k\}$. Thus, no more than $1,000$ parcels can be processed at a chute in $30,000$ seconds, and hence, the maximum possible load for the operation for $k=30$ chutes turns out to be $30,000$. We assume a projected load of $29,335$ parcels (close to peak capacity). The planning algorithm is able to process all parcels in the case of unrestricted layouts; and $29,148$ parcels in the case of restricted layouts. 


To compare the performance of our algorithm, we use a greedy heuristic (GREEDY) as a baseline. This heuristic pushes a parcel $P_m$ into the first free chute $C_j$ such that $Q_{mj}=1$; i.e., a parcel is pushed into the first free chute if it can be allocated to that chute. We compare the performance of the two algorithms based on the following key performance indicators: (i) \emph{Number of re-circulations $Rc$} (Number of parcels circulated more than once before entering a chute/rejection chute), (ii) \emph{Number of rejections $Rj$} (total number of parcels entering the rejection chute in a wave), and (iii) \emph{Average Sorting time $S_t$} (time elapsed between the point at which a parcel crosses the OCR reader (enters the system) and the point when it enters a chute/rejection chute, averaged over all parcels in a wave).

\begin{figure}[tb]
         \centering
         \includegraphics[width=0.52\textwidth]{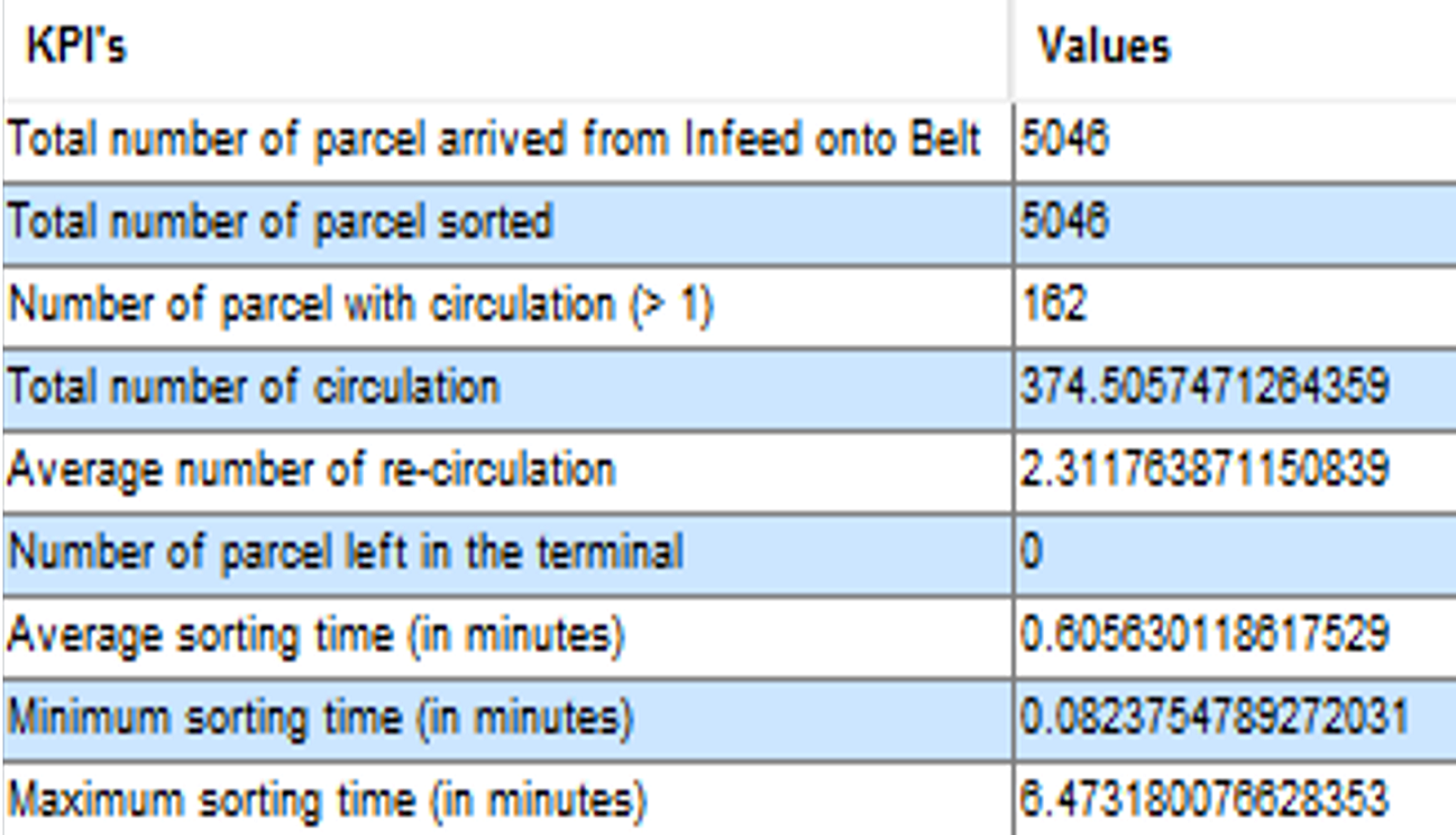}
         \caption{KPIs for GREEDY in unrestricted layout, captured in simulator interface.}
         \label{fig:unckpi2}
\end{figure}

We ran several experiments with varying load capacities and profiles for waves e.g., loads with \emph{benign} capacity ($\approx 50\%$ capacity); waves with load-profiles (i.e., parcel-destination matching) similar to that of the forecast, and random load-profiles. In the case of waves with benign load capacity, both GREEDY and OPTSORT performed equally well with zero rejections and re-circulations (with slight variations in average sorting times). On the other hand, once can observe a marked difference in KPIs once the load capacity increases. To demonstrate this, we present the results of one set of such experiments here. Two waves were sent in comprising of $N=2,523$ parcels ($\approx 85\%$ of max. capacity) each, with random $\tau_m$ for $m\in\{1,\dots,2523\}$. Figure~\ref{fig:unckpi2} shows the KPIs for GREEDY with no layout restrictions. Other figures are not presented due to space constraints.

The KPIs for GREEDY with no layout restrictions are $Rc=Rj=162$ and $S_t=0.6$ min; while the same for OPTSORT with $\bar{\mathcal{C}}_j=50$, they are $Rc=0$, $Rj=190$, and $S_t=0.499$ min. Thus, OPTSORT results in no re-circulation with a much lower average sorting time. However, the number of rejections for OPTSORT is higher than GREEDY. This is due to the fact that $\bar{\mathcal{C}}_j=50$ turns out to be too conservative. Increasing $\bar{\mathcal{C}}_j$ to $55$, removes the rejections while maintaining zero re-circulations and smaller average sorting time. The KPIs for unrestricted and restricted layout scenarios are captured in Table~\ref{tab:output}. The results for the Direct-chute+restricted layout scenario are almost similar and are omitted due to space constraints.

\begin{table}[t]
\centering
 \begin{tabular}{| c | c | c | c |} 
 \hline
 Algorithm+Situation & Rc & Rj & $S_t(min)$\\
 \hline
 Unrestricted GREEDY & 162 & 162 & 0.606 \\
 \hline
 Unrestricted OPTSORT ($\bar{\mathcal{C}}_j=\mathcal{C}_j=50$) & $\mathbf{0}$ & 190 & 0.499 \\
 \hline
 Unrestricted OPTSORT ($\bar{\mathcal{C}}_j=55$) & $\mathbf{0}$ & $\mathbf{0}$ & $\mathbf{0.495}$ \\
 \hline
 Restricted GREEDY & 55 & 55 & 0.557 \\
 \hline
 Restricted OPTSORT ($\bar{\mathcal{C}}_j=\mathcal{C}_j=50$) & $\mathbf{0}$ & 341 & 0.507  \\
 \hline
 Restricted OPTSORT ($\bar{\mathcal{C}}_j=65$) & $\mathbf{0}$ & $\mathbf{0}$ & $\mathbf{0.491}$  \\ 
 \hline
 \end{tabular}
 \caption{Key Performance Indicators for various test situations.}
 \label{tab:output}
\end{table} 

Experiments with varying human efficiency/productivity (randomly between 0.8 and 1.2) were also simulated using the digital twin. Variations in human efficiency lead to variations in processing time per parcel and consequently, the overall throughput of the system may change. Our objective was to study the robustness of OPTSORT against such variations and its iterative improvement via feedback from digital twin. For instance, for the unrestricted layout scenario, with $\bar{\mathcal{C}}_j=60$, no parcels were rejected or re-circulated with varying human efficiency (as opposed to $\bar{\mathcal{C}}_j=55$ with nominal human efficiencies). However, there is one caveat to this study. OPTSORT assumes that the efficiency of person manning a chute is known \emph{a priori}; i.e., it can be different from nominal ($100\%$); but it needs to be known. Such an assumption may not be feasible in a real-world scenario; and work is in progress to reconcile this difference.

\section{CONCLUSIONS}

We studied how a combination of optimization and 
high-fidelity simulatable digital twin can be used to improve the operations of a sorting center terminal. We developed a two-stage optimization procedure OPTSORT intended to maximize throughput of a sorting terminal, validated and refined using feedback from a simulatable digital twin of the sorting terminal. We showed that the digital twin is a critical component, helping to balance performance with robustness. Future work in this regard is threefold: first, to model more real-world constraints and scenarios such as uncertain human efficiency and time delays in processing; second, use efficient heuristics instead of MILP for faster processing; and finally, integrate various logistics operations such as vehicle routing with sorting logic to solve large-scale enterprise problems. 


\footnotesize

\bibliographystyle{wsc}

\bibliography{refs}

\section*{AUTHOR BIOGRAPHIES}


\noindent {\bf SUPRATIM GHOSH} is a scientist with TCS Research. He earned his Ph.D and M. S. in Electrical Engineering, and M. A. in Mathematics from the Pennsylvania State University. His research interests include applications of control and optimization techniques to supply chain and robotics. His email address is \email{supratim.ghosh2@tcs.com}. \\

\noindent{\bf ARITRA PAL} is a researcher with TCS Research. He earned his M. Tech in Mathematics and Computing from Indian Institute of Technology, Patna. His research interests include probability, statistics and optimization. His email address is \email{p.aritra@tcs.com}. \\

\noindent{{\bf PRASHANT KUMAR} is a Researcher at TCS Research. He holds a B.Tech in Information Technology from the BIT Sindri. His research interests are in the area of actor based simulations. His email address is \email{kumar.prashant10@tcs.com}}.
 \\

\noindent{\bf ANKUSH OJHA} is a researcher with TCS Research. He earned his M. Tech degree in Industrial and Management Engineering from Indian Institute of Technology, Kanpur. His research interests include application of reinforcement learning and optimization to problems in supply chain and transportation. His email address is  \email{ojha.ankush@tcs.com}. \\

\noindent {\bf ADITYA A. PARANJAPE} is currently a  Lecturer in the Department of Aeronautics at Imperial College London, and on sabbatical from TCS Research where he holds the position of Scientist in Software Systems and Services. He earned his Ph.D. in Aerospace Engineering from the University of Illinois at Urbana-Champaign in 2011. His research interests are centered around control theory, multi-agent systems, and reinforcement learning.  His email address is \email{aditya.paranjape@tcs.com}. \\

\noindent {{\bf SOUVIK BARAT} is a  Principal Scientist at Tata Consultancy Services Research. He earned his Ph.D. in Computer Science from Middlesex University. His research interests include simulation, agent
and actor technology, enterprise modeling and digital twin. His email address is \email{souvik.barat@tcs.com}.}. \\

\noindent {{\bf Harshad Khadilkar} is a senior scientist with TCS Research. He holds a PhD in Aeronautics and Astronautics from the Massachusetts Institute of Technology. His research interests are in the area of optimal decision making for industrial operations. His email ID is \email{harshad.khadilkar@tcs.com}. 

\end{document}